\documentclass[pre,superscriptaddress,showkeys,showpacs,twocolumn]{revtex4-1}
\usepackage{graphicx}
\usepackage{latexsym}
\usepackage{amsmath}
\begin {document}
\title {Bipartitioning of directed and mixed random graphs}
\author{Adam Lipowski}
\affiliation{Faculty of Physics, Adam Mickiewicz University, Pozna\'{n}, Poland}
\author{Ant\'onio  Luis Ferreira}
\affiliation{Departamento de F\'{i}sica, I3N, Universidade de Aveiro,  Portugal}
\author{Dorota Lipowska}
\affiliation{Faculty of Modern Languages and Literature, Adam Mickiewicz University, Pozna\'{n}, Poland}
\author{Manuel A. Barroso}
\affiliation{Departamento de F\'{i}sica, I3N, Universidade de Aveiro,  Portugal}
\begin {abstract} We show that an intricate relation of cluster properties and optimal bipartitions, which takes place in undirected random graphs, extends to directed and mixed random graphs. In particular, the satisfability threshold coincides with the relative size of the giant OUT component reaching~{1/2}. Moreover, when  counting undirected links as two directed ones, the partition cost, and cluster properties, as well as location of the replica symmetry breaking transition for these random graphs depend primarily on the total number of directed links and not on their specific distribution.

\end{abstract}

\maketitle

\section{Introduction}
Statistical mechanics methodology is frequently used in the studies of various optimization problems~\cite{hartmann2006}. 
Presence of  quenched disorder, energy barriers, or various phase transitions in such problems implies interesting analogies to some glassy or magnetic  systems and the usage of methods developed in the physical sciences turns out to be remarkably successful~\cite{krzakala2016}.

A graph bipartitioning is an optimization problem that appears in various contexts such as  VLSI circuit design~\cite{karypis}, parallel computing~\cite{pothen} or computer vision~\cite{kolmogorov}.    Statistical mechanics approaches are  particularly fruitful in the undirected random graph version of this problem. Such  a version was studied numerically using a simulated annealing~\cite{banavar,martin} or an extremal optimization~\cite{boetcher} but important analytical results were also obtained using  a replica method~\cite{fu,liao,mezard}, the technique that was primarily developed for studying disordered systems. In a more recent  work, in which the structure of nearly optimal partitions was analyzed, some predictions concerning the replica symmetry breaking in this problem were made~\cite{percus}  that were subsequently verified using the belief propagation method~\cite{zdeborova}.

In contrast to undirected random graphs, bipartitioning problem in other classes of graphs is less understood. Taking into account that in most of systems links are only approximately symmetric, it would be desirable to examine partitioning problem on random graphs with directed links, and that was our main motivation. Let us notice that despite an apparently simple modification, directed graphs are much more difficult to analyze. For example, the asymmetricity of  adjacency matrix and of  Laplacian considerably complicates  their spectral analysis~\cite{chung}. Similar complications affect also a closely related problem of community detection in directed graphs~\cite{malliaros}. 

\section{model}

\begin{figure}
\includegraphics[width=\columnwidth]{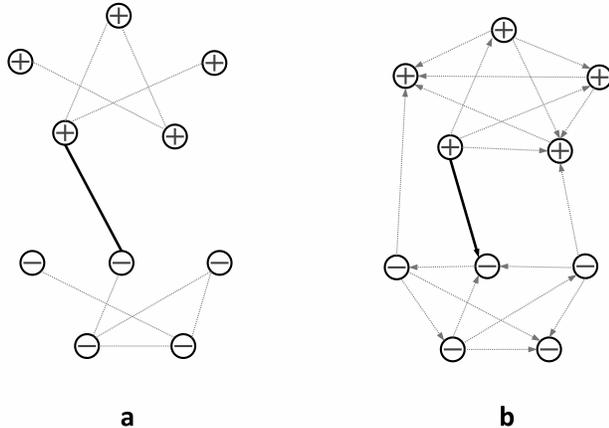}
\vspace{-10mm}
\caption{In the graph bipartition problem, one has to separate nodes into two classes, $\oplus$~and~$\ominus$,  so that the partitioning cost is minimal. (a) For the undirected graphs, the partition cost equals to the number of links that join nodes with opposite signs (thick line).  (b) For directed graphs, only links from~$\oplus$ to~$\ominus$ contribute to the partition cost.}
\label{graph}
\end{figure}

In the graph bipartitioning, one has to divide $N$~vertices into two classes of  equal size, here marked as~$\oplus$ and~$\ominus$, so that the partition cost is minimal. For undirected graphs, the partition cost is equal to the number of links with vertices of opposite signs (Fig.~\ref{graph}a). In the present paper, we examine an extension of this problem to directed or mixed Erd\"{o}s-R\'enyi graphs~\cite{erdos}, where both directed and undirected links are present. To construct such graphs, we place $pN$~links among randomly selected pairs of vertices. With probabilities $p_{u}$ and 1-$p_{u}$, the link is undirected or directed, respectively, and in the latter case its orientation is chosen randomly.  As limiting cases, in such a way we can generate undirected ($p_{u}=1$) or directed ($p_{u}=0$) random graphs. 

Bipartitioning of undirected graphs bears some analogy to the Ising model, in which on each vertex $i$, there is a spin variable $S_i=\pm 1$, and the system is described with the following Hamiltonian
\begin{equation}
H=-\sum_{(i,j)} S_iS_j,
\label{hamiltonian}
\end{equation}
In the above equation, summation is over pairs of vertices connected by a link, and the system is subject to the constraint that the numbers of~$\oplus$ and~$\ominus$ are equal, namely $\sum_{i=1}^N S_i=0$.
In terms of spin variables, the normalized partition cost can be written as
\begin{equation}
b=\frac{1}{2N}\sum_{(i,j)} (1-S_iS_j),
\label{cost1}
\end{equation}
Finding an optimal partition becomes thus equivalent to finding the lowest energy of the Ising model subject to the constraint of zero magnetization.
A number of approaches to the graph bipartitioning were developed, which exploit the above analogy with the Ising model. However, for directed or mixed graphs, the  analogy becomes more intricate. One can retain spin variables $S_i$, but the spin dynamics looses a detailed balance and a description in terms of the Hamiltonian like (\ref{hamiltonian}) is less obvious~\cite{godreche,sanchez}. 
Neglecting the fact that partitioning of directed graphs involves a number of mathematical subtleties~\cite{charikar,feige}, we use the partition cost  that counts only links of which the origin is~$\oplus$ and the end is~$\ominus$ (Fig.~\ref{graph}b). With such a definition, the partition cost for directed graphs in terms of spin variables might be written as
\begin{equation}
b=\frac{1}{N}\sum_{(i,j)} \delta_{S_i,1}\delta_{S_j,-1},
\label{cost_directed}
\end{equation}
where summation is over directed links that originate at $i$ and end at $j$.
Actually,  definition (\ref{cost_directed}) can be applied also for undirected graphs and is equivalent to  (\ref{cost1}) since an undirected link could be considered as composed of two directed and opposite links, and if it joins verticies of different signs, it will be counted but only once.


The cost function as defined in Eq.(\ref{cost_directed}) was used for example in some studies on the complexity of the partition algorithms \cite{feige}. Let us notice, that in principle for directed graphs we can use the symmetric cost function (\ref{cost1}). In such a way, however, directedness of the graph is lost since the problem becomes basically equivalent to the undirected case. The only difference would come from the possibly existing pairs of vertices joined via two oppositely oriented links. For random graphs with finite $p$ such pairs of vertices appear with a negligably small probability.

\section{cluster properties}
Partition of random graphs relates to their percolative behavior. In particular, for undirected random graphs, it is known that as long as the size of the giant cluster is smaller than $N/2$, one typically finds partitions with $b=0$~\cite{luczak,mezard}. Indeed, if the entire giant cluster is set as (for example)~$\oplus$, then the remaining~$\oplus$  and  $N/2$~$\ominus$ most likely can be distributed among small isolated clusters, so that the partition cost is zero. Since the relative size of the giant cluster $g$ for random undirected graphs ($p_u=1$) obeys the equation
\begin{equation}
g=1-\exp{(-2pg)},
\label{giant}
\end{equation}
the  satisfiability regime with $b=0$ persists in undirected graphs up to  $p=p_s^u=\ln 2$. Let us note that distributing $pN$ links among $N$ vertices, we generate a random graph of the average vertex degree $2p$ and hence $p=p_c^u=1/2$ marks the percolation threshold in undirected graphs.

For $p>p_s^u=\ln 2$, the giant cluster is greater than $N/2$ and some of its vertices must be~$\ominus$, which implies a positive partition cost:  $b>0$. When $p$ is not much greater than $1/2$, optimal or nearly optimal partitions might be constructed based on the cluster structure. In particular, an efficient strategy is to turn into~$\ominus$  some tree-like decorations of the giant component~\cite{percus,benjamini}.

Such a behavior  of undirected graphs prompts to search for a similar scenario in directed and mixed graphs.  Let us notice, however, that percolative properties of such graphs are more subtle. Directedness of links implies that clusters are now defined as collections of vertices that might be reached from a given vertex (OUT components) or of those from which a given vertex can be reached (IN components). One can also distinguish  strongly connected components, of which each vertex can be reached from any other vertex that belongs to the component, as well as the so-called tendrils or tubes~\cite{bowtie,timar}. The satisfiability threshold for undirected graphs $p_s^u=\ln 2$ results from the condition that the largest cluster that can be uniformly magnetized reaches $N/2$. It seems that in the presence of directed links, the largest OUT component magnetized as~$\oplus$  might play such role. Indeed, any vertex that does not belong to the OUT component might be connected to it but only via incoming (to the OUT component) link. It will not increase the partition cost even if such vertex is~$\ominus$. The above strategy would require some modifications when the average sizes of IN and OUT components are different, but for random graphs analyzed in the present paper, this is not the case. Moreover, since the partition cost (\ref{cost_directed}) is invariant with respect to simultaneous flipping of spins and link directions, we can equally well expect that the satisfiability threshold appears when the IN component that is magnetized as~$\ominus$ reaches $N/2$.

For directed graphs ($p_u=0$), the size of OUT and IN components  can be calculated using generating function approach\cite{dorog}. Placing $pN$ directed links is then equivalent to the following method: for each directed pair of vertices $(i,j)$, place a link from $i$ to $j$ with probability $p/N$. For such a case, an explicit calculation of generating function is straightforward~\cite{lipgont} and one obtains that the average relative size of the largest OUT component $g_O$ satisfies the equation
\begin{equation}
g_O=1-\exp{(-pg_O)}.
\label{giant_out}
\end{equation}

Let  us notice that the above equation, except for the factor~2, is the same as Eq.~(\ref{giant}), which implies that for directed graphs the percolation transition takes place at $p=p_c^d=1$ and $g_O=1/2$  at $p=p_s^d=2\ln 2$. More generally, the size of the largest OUT component for a directed graph  at a given $p$ (above a percolation threshold $p_c^d$) is exactly the same as the size of the giant cluster for undirected graph at twice smaller~$p$.

\begin{figure}
\includegraphics[width=\columnwidth]{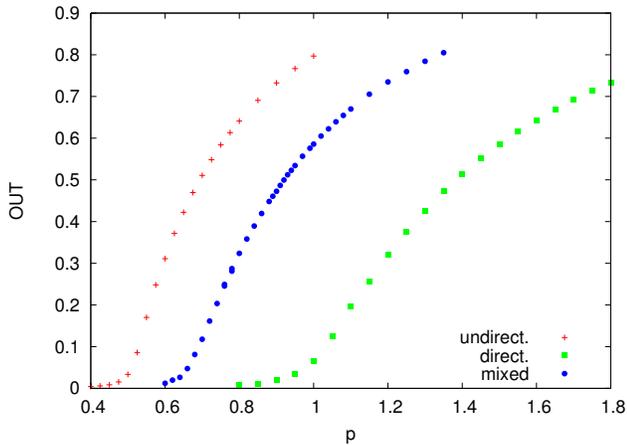}
\vspace{-7mm}
\caption{The average relative size of the giant  OUT components as a function of~$p$. Simulations were made for $N=3\cdot 10^4$ and for each~$p$ we generated 100 graphs. Close to the percolation transition the numerical data exibit typical finite-size rounding.}
\label{percol0}
\end{figure}

For mixed graphs, analytical calculations seem to be less straightforward and we resort to numerical calculations. We calculated the largest OUT component for mixed graphs with $p_u=0.5$,  but in Fig.~\ref{percol0} we present also the results for undirected and directed graphs (for undirected graphs, the OUT component is the same as the giant cluster). Numerical results agree with the expectation that for undirected graphs, the percolation transition takes place at $p_c^u=1/2$ and for directed ones at $p_c^d=1$. For mixed random graphs, some heuristic arguments suggest that two vertices  connected with a directed link with probabilities 1/2 might be considered as connected via an undirected link or as disconnected~\cite{herrmann}. 
It means that a mixed random graph having $pN$ links, with $p(1-p_u)N$ of them  directed, is equivalent to the random graph with  $pp_uN+p(1-p_u)N/2=p(1+p_u)N/2$ undirected links. One can thus expect a percolation transition taking place at $p=p_c^m$ that obeys the equation
\begin{equation}
p_c^m(1+p_u)=1,
\label{pc}
\end{equation}
 which for $p_u=0.5$ gives $p_c^m=2/3$, a value which is consistent with numerical simulations (Fig.~\ref{percol0}). 

\begin{figure}
\includegraphics[width=\columnwidth]{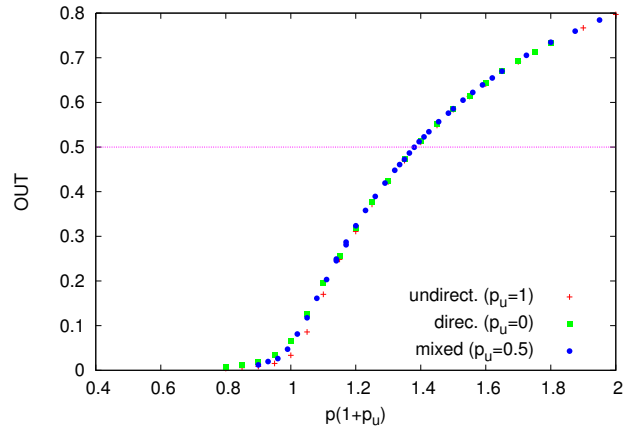}
\vspace{-10mm}
\caption{The numerical results of Fig.~\ref{percol0} plotted as a function of $p(1+p_u)$. Let us notice that $p(1+p_u)N$ equals to the total number of directed links, provided that we count an undirected link as two directed links.}
\label{percol}
\end{figure}

It follows from Eq.~(\ref{pc}) that when plotted as a function of $p(1+p_u)$ rather than~$p$, our numerical results should all exhibit a percolation transition at the same point. Moreover, according to Eqs. (\ref{giant}) and~(\ref{giant_out}),  the size of OUT components for directed and undirected graphs should be the same for any $p(1+p_u)$. In Fig.~\ref{percol}, we present the size of OUT components as a function of $p(1+p_u)$. As expected, the three kinds of graphs have the percolation transition at $p(1+p_u)=1$, but what is more, not only the data for directed and undirected graphs collapse on a single curve but also the data for mixed graphs collapse onto this curve as well. It suggests that the OUT component satisfies (or nearly satisfies) an equation similar to Eq.~(\ref{giant_out})  also for mixed graphs, with~$p$ replaced by $p(1+p_u)$. We will not analyze this issue further but we hope that a suitable extension of the generating function approach could explain these numerical findings.

\section{simulated annealing}
Having examined the percolative behavior of directed and mixed random graphs, we can develop a method of their partitioning. Similarly to the undirected graphs, we expect that as long as the relative size of the OUT component $g_O$ is smaller than 1/2, we may set it as~$\oplus$ and hope that the rest of vertices will be arranged with a zero partition cost. More challanging is the case of $g_O>1/2$. In such a case, a uniform magnetization of the OUT component is excluded and some of its vertices must be set as~$\ominus$, which will inevitably generate a partition cost. To make it optimal for undirected graphs, Percus {\em et al}.~\cite{percus} developed a combinatorial analysis of some outer tree-like decorations of a giant component. 

Taking into account more complex structure of directed and mixed graphs, we gave up analytical manipulations and resort to a simulated annealing, which takes into account the structure of a cluster. More precisely, for a given graph, we find the largest OUT component, set it as~$\oplus$, and classify its vertices according to the distance from its boundary. Then, depending on the size of the OUT component (we assume that $g_O>1/2$), we set the required number of vertices as~$\ominus$. We choose them taking into account their distance from the boundary, namely, vertices close to the boundary are preferably set as~$\ominus$---similarly as decorations of undirected graphs of Percus {\em et al.}~\cite{percus}. Thus we expect that such a procedure will allow for an insertion of a relatively large number of~$\ominus$ at low cost. The configuration determined in this way becomes the initial one, for which the simulated annealing is used to reshuffle~$\ominus$ in the OUT component so that the minimal cost is reached.  The annealing algorithm selects a pair of oppositely magnetized vertices and exchanges them according to the Metropolis update that accepts the partition-cost increase $\Delta b$  with probability $\exp(-\Delta b/T)$. During the run, the temperature~$T$ is reduced as $T=T_0\exp(-rt)$, where $r$ is the cooling rate and $t$ is the simulation time (unit of time is defined as an update of $N$ pairs of vertices). 
Because we expect that the optimal configuration to some extent will resemble the initial one, we do not want to destroy its structure during the run especially at the high-temperature regime. That is why the high-temperature regime should not last  too long nor the initial temperature should not be too large. We found that the initial temperature $T_0=100$, and the cooling rate $r=10^{-5}$ usually lead to satisfactory partitions and numerical results presented in the following are obtained for such a choice of parameters.

In the regime with $g_O<1/2$, a slightly different simulations were made. We find the OUT component, set it as~$\oplus$ and the rest of~$\oplus$ as well as $N/2$ of~$\ominus$ are distributed randomly. Then we run the simulated annealing that reshuffles  spins only on vertices that do not belong to the OUT component (which is kept unchanged). As expected, in the regime $g_O<1/2$ partitions with $b=0$ (or with very small~$b$) are easily found.  

\subsection{partition cost}
\begin{figure}
\includegraphics[width=\columnwidth]{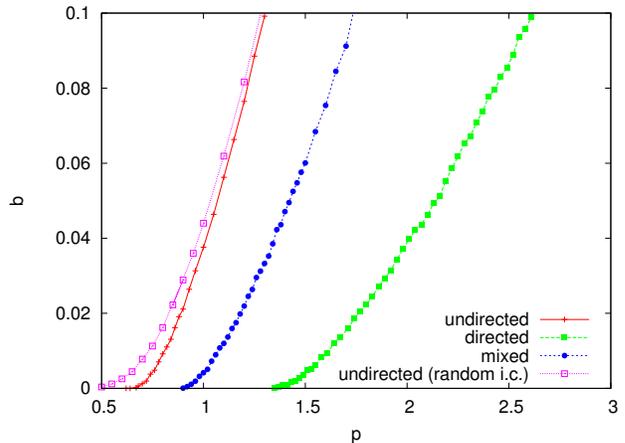}
\vspace{-8mm}
\caption{The average partition cost $b$ as a function of the link probability~$p$ for undirected, directed and mixed ($p_u=0.5$) random graphs ($N=3000$). Results are obtained using simulated annealing with an initial configuration taking into account the cluster structure. For undirected graphs, the simulated annealing that neglects the cluster structure and uses random initial configurations (random i.c.) gives noticably worse results.}
\label{cost}
\end{figure}

\begin{figure}
\includegraphics[width=\columnwidth]{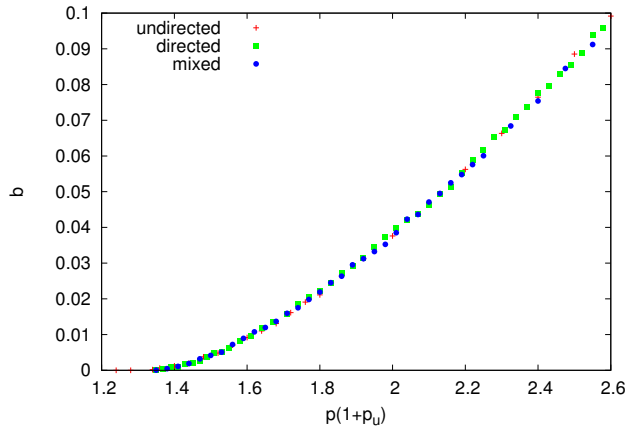}
\vspace{-8mm}
\caption{The average partition cost $b$ as a function of the rescaled link probability $p(1+p_u)$. Considering an undirected link as two directed ones, such a collapse shows that it is the total number of directed links that determines the cost.}
\label{cost-collapse}
\end{figure}

For undirected graphs, our numerical results (Fig.~\ref{cost}) reproduce the already known  results and show that the partition cost becomes positive for $p>p_s^u\ln 2\approx 0.693$. Analogous results are obtained for directed and mixed graphs, and the emergence  of a positve partition cost coincides with  $g_O=1/2$, namely, it takes palce at $p=p_s^d=2\ln 2\approx 1.386$ for directed graphs and $p=p_s^m=1.5\ln 2\approx 1.039$ for mixed graphs. What is rather surprising for us is the overlap of the cost-function data when plotted as a function of $p(1+p_u)$ (Fig.~\ref{cost-collapse}). Similarly to the behavior of the size of the OUT component, such a collapse indicates that the partition cost depends solely on the total number of directed bonds in the system and not on their specific distribution.

\subsection{replica symmetry}
An interesting aspect of graph partitioning is related to the nature of the solution space and a possible breakdown of the so-called replica symmetry. A predicition was made by Percus {\it et al.} that in the partitioning problem of undirected random graphs, the replica symmetry breaking should occur inside the $b>0$ regime, while in most other optimization problems it takes place inside the costless phase~\cite{percus}. Recently, using a belief propagation method, it was shown that the replica symmetry breaking occurs for $p>p_r^u\approx 0.736$, in agreement with the Percus {\it et al.} conjecture~\cite{zdeborova}. 

The replica symmetry is related to the similarity of different ground state configurations. In a replica symmetric phase, such configurations are to a large extent similar, while in a replica symmetry-broken phase, they are much different. Such a symmetry was intensively studied in the hope to clarify the nature of  ordering in spin glasses~\cite{marinari,young} as well as in various optimization contexts~\cite{krzakala,monasson}. To examine this symmetry, we generate a graph and run the simulated annealing for two different replicas A and~B. Assuming that at the end of the run these replicas are specified by their spin configurations $\{S_i^A\}$ and $\{S_i^B\}$, respectively, we calculate the overlap~$q$ defined as follows
\begin{equation}
q=1/N\sum_{i=1}^N S_i^AS_i^B .
\label{overlap}
\end{equation}
Actually, since a graph structure to some extent determines the initial configuration, the replicas  differ only in the distribution of~$\ominus$ that have to be placed in the OUT component. For a given graph, specified by~$p$ and~$p_u$, to calculate~$q$ we average over $10^2$ pairs of replicas and  we also average over $10^2$ different graphs.

On general grounds, one expects that in the replica symmetric phase, the distribution of~$q$ is strongly peaked at a value close to $q=1$, which corresponds to a single-valley structure of the ground state. In the replica broken-symmetry phase, much broader distribution is expected, which in some systems even at $q=0$ remains positive.

\begin{figure}
\includegraphics[width=\columnwidth]{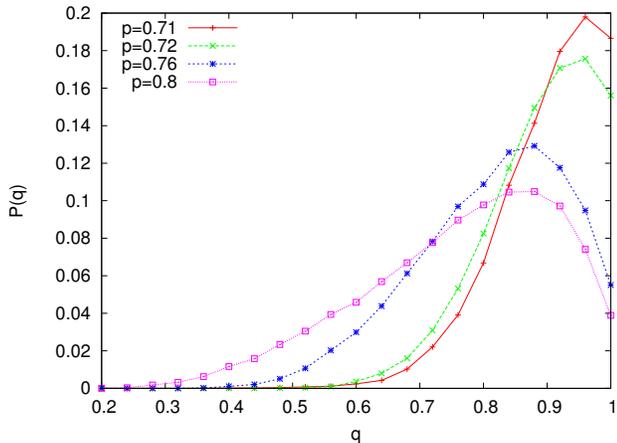}
\vspace{-8mm}
\caption{The overlap probability distribution $P(q)$ calculated for undirected random graphs ($N=500$).}
\label{bin-undir}
\end{figure}

\begin{figure}
\includegraphics[trim=0cm 0cm 6cm 0cm, clip=true, width=\columnwidth]{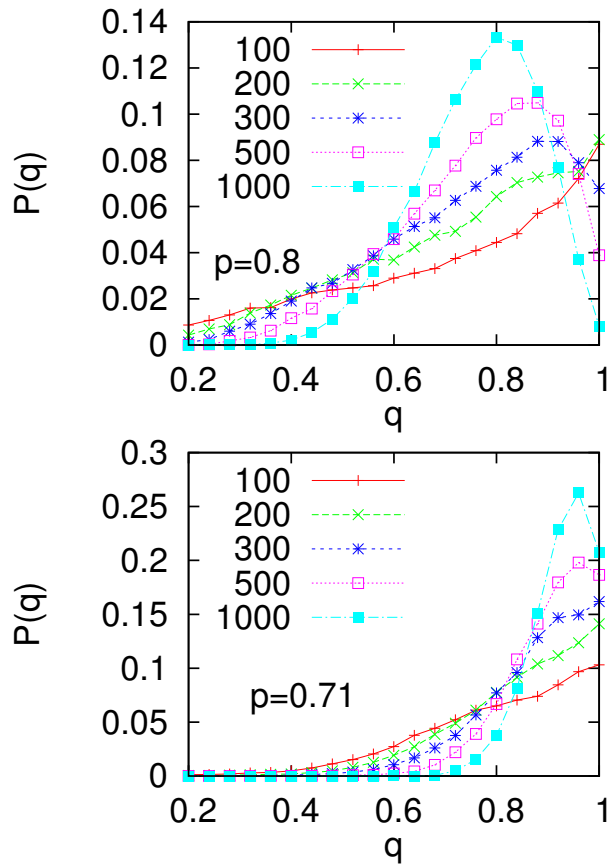}
\vspace{-8mm}
\caption{The overlap probability distribution $P(q)$ calculated for undirected random graphs and $p=0.8$  and 0.71. The size of the system $N$ is shown in the legend.}
\label{size-undir}
\end{figure}

The distribution of~$q$ in our simulations for undirected random graphs (Fig.~\ref{bin-undir}) clearly shows two regimes:  for $p=0.71$ and~0.72, the distribution is strongly peaked around $q=1$, while for larger values $p=0.76$ and~0.8, the peak is smaller and tails at small~$q$ are much heavier.  The dependence on  the system size $N$ (Fig.\ref{size-undir}) suggests that such a difference in the behaviour of $P(q)$ is likely to persist also for larger $N$. 


Let us notice that our algorithm starts from configurations that are only partially random and thus even in the replica broken-symmetry phase, the replicas are not independent. Consequently, we cannot expect that $P(q)$ will remain positive at $q=0$. In our opinion, the numerical results support, albeit with a smaller precision, the previous estimation $p_r^u=0.736$ as a replica symmetry breaking transition~\cite{zdeborova}.

\begin{figure} 
\includegraphics[width=\columnwidth]{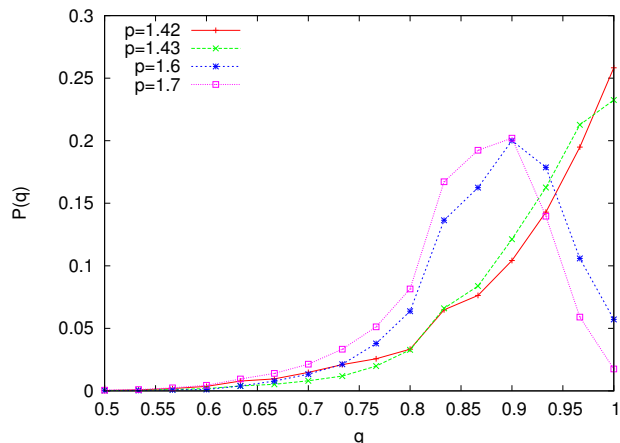}
\vspace{-8mm}
\caption{The overlap probability distribution $P(q)$ calculated for directed random graphs ($N=500$).}
\label{bin-direct}
\end{figure}

\begin{figure}
\includegraphics[trim=0cm 0cm 6cm 0cm, clip=true, width=\columnwidth]{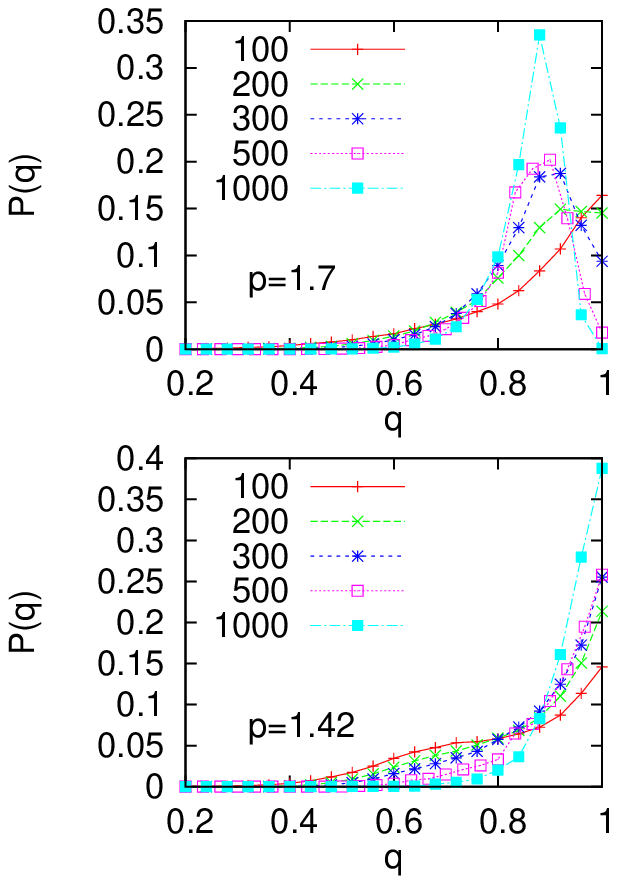}
\vspace{-8mm}
\caption{The overlap probability distribution $P(q)$ calculated for directed random graphs and $p=1.7$  and 1.42. The size of the system $N$ is shown in the legend.}
\label{size-dir}
\end{figure}

A similar behavior is observed for directed graphs (Fig.~\ref{bin-direct}-\ref{size-dir}) and we estimate that the transition takes place at $1.43 <p_r^d<1.6$, and this is nearly twice the value of $p_r^u$ for undirected graphs. Less convincing are the results for mixed graphs but the regime with a fast decay at small~$q$ ($p=1.06$, 1.08) and a slow decay ($p=1.15$, 1.2) can be also distinguished. Rescaling the undirected graph transition at $p=0.736$ \cite{zdeborova} with the factor $(1+p_u)=1.5$, we obtain $p_r^m=1.5p_r^u=1.104$, which clearly falls within the range (1.08, 1.15). It indicates that when expressed in terms of $p(1+p_u)$, replica symmetry breaking  transitions for undirected, directed, and mixed random graphs  take place at (nearly) the same value. Similarly to the percolation transitions, in the examined class of random graphs, replica symmetry breaking transitions  seem to depend only on the total number of directed links in the graph.

\begin{figure}
\includegraphics[width=\columnwidth]{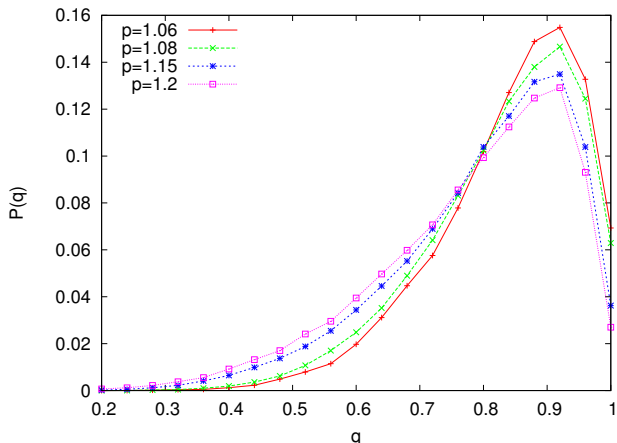}
\vspace{-8mm}
\caption{The overlap probability distribution $P(q)$ calculated for mixed random graphs ($N=500$).}
\label{bin-mixed}
\end{figure}

The system size $N=500$ used in the calculation of overlap $P(q)$ seems to be sufficiently large to detect the different regimes and sufficiently small to allow extensive averaging. For larger $N$ we expect that peaks in the replica symmetric regime get sharper that will indicate a genuine phase transition. Of course, more detailed analysis of finite size effects would be desirable but that would require much longer simulations. Let us notice that for spin glasses the overlap $P(q)$ is often calculated using rather small systems \cite{katz}.

\section{conclusions}
In undirected random graphs when the size of the giant cluster is smaller than half of the system, the zero-cost bipartitions usually could be found. The idea is to keep the giant cluster uniformly magnetized and then the rest of vertices can usually be marked without any partition cost.
When the number of links increases and the size of the giant cluster exceeds, but not much, half of the system,  the partition cost is unavoidable, however, the optimal partitions still contain the footprint of the cluster structure. It means that for a given graph, they are all similar or, in other words, the system preserves the replica symmetry. Upon a further increase in the number of links, the relation of optimal partitions to the giant cluster weakens and  the replica symmetry gets broken.

As our main result, we show that to a large extent the above scenario takes place also in directed and mixed random graphs with a giant component replaced by the giant OUT component. What is more, however, the partition cost, the satisfability threshold, and the replica symmetry breaking transition   in undirected, directed, and mixed random graphs seem to depend only on the total number of directed links in the graph (counting undirected link as two directed ones).  Our simulations show a similar behavior of cluster properties of these graphs, where the percolation threshold and  the size of the giant component exhibit analogous dependence on the total number of directed links. A simple idea that in percolation problems, an undirected link might be equivalent to  two directed links finds a strong support in some models on regular lattices~\cite{herrmann}. Our work shows that it extends also to some partitioning problems.

Acknowledgements: 
This work was partially funded by FEDER funds through the COMPETE 2020 Programme and National Funds throught FCT - Portuguese Foundation for Science and Technology under the project UID/CTM/50025/2013. 

\end {document}